\documentclass[sigconf]{acmart}
\renewcommand\footnotetextcopyrightpermission[1]{}
\AtBeginDocument{%
  }

\setcopyright{none}   

\copyrightyear{2026}
\acmYear{2026}

\acmDOI{}             
\acmISBN{}            

\acmConference[UKAIRS '26]
{UK AI Research Symposium (UKAIRS 2026)}
{November 24--25, 2026}
{United Kingdom}

\begin{document}
\title{A Dual-Edge Spatial–Jacobian Image Graph for Interpretable Diabetic Retinopathy Grading}

\author{Inam Ullah}
\email{i1n23@soton.ac.uk}
\orcid{https://orcid.org/0000-0001-9488-035X}
\affiliation{%
  \institution{University of Southampton}
  \city{Southampton}
  \country{United Kingdom}
}

\author{Imran Razzak}
\affiliation{%
  \institution{Mohamed bin Zayed University of Artificial Intelligence}
  \city{Abu Dhabi}
  \country{United Arab Emirates}
  \email{Imran.Razzak@mbzuai.ac.ae}}





\author{Shoaib Jameel}
\affiliation{%
  \institution{University of Southampton}
  \city{Southampton}
  \country{United Kingdom}}
\email{M.S.Jameel@southampton.ac.uk}


\begin{abstract}
Automated diabetic retinopathy (DR) grading from colour fundus photographs can achieve strong predictive performance, but clinical interpretation requires more than an image-level label. It requires understanding how lesion evidence is distributed around retinal vessels and how this evidence relates to quantitative vascular biomarkers. We present a dual-edge spatial--Jacobian image graph for interpretable DR grading. Each fundus image is represented as a graph node with four aligned evidence streams: AutoMorph vessel information ($X_1$), DR-XAI-style lesion evidence maps ($X_2$), a 128-dimensional lesion-based contrastive image embedding ($X_3$), and AutoMorph morphometric biomarkers ($X_4$). The spatial edge branch ($X_{12}$) encodes vessel--lesion geometry, while the Jacobian branch ($X_{34}$) models embedding--biomarker sensitivity. Lightweight two-token attention fuses both edge families into a final image graph. On 2,910 matched non-augmented APTOS images, the full graph achieves 0.8076 accuracy, 0.8312 quadratic weighted kappa, 0.5915 macro-F1, and 0.9330 adjacent-grade accuracy; referable DR reaches 0.9055 accuracy and 0.9711 AUROC. The framework is positioned as an explainable representation-learning tool for lesion--biomarker hypothesis generation, rather than as a deployment-ready clinical classifier. The code is available at \url{https://github.com/Inamullah-Colab/dual-edge-dr-graph-xai}.
\end{abstract}

\maketitle

\begin{figure}[t]
\centering
\includegraphics[width=0.48\textwidth]{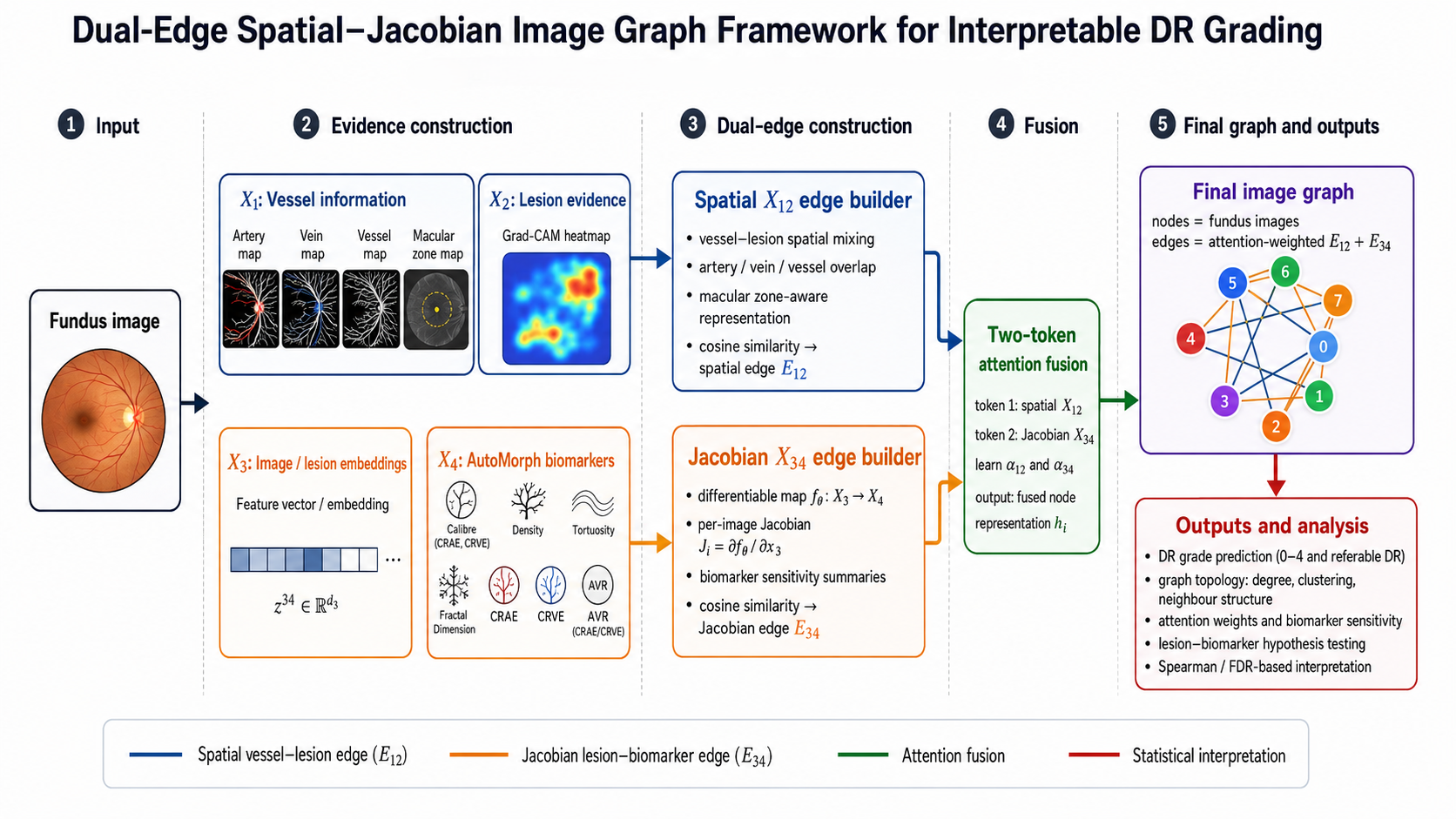}
\caption{Dual-edge spatial--Jacobian graph workflow. Fundus
images are represented by four aligned evidence streams. Vessel maps
and lesion evidence form the spatial ~$X_{12}$ edge branch, while
lesion-based embeddings and AutoMorph biomarkers form the Jacobian
~$X_{34}$ edge branch. Two-token attention fuses both branches into a
final image graph for DR grading and interpretation.}
\label{fig:workflow}
\end{figure}

\section{Introduction}

Diabetic retinopathy is a retinal microvascular complication of diabetes and remains a major cause of preventable visual impairment. Deep learning systems have shown that referable DR can be detected from fundus photographs with high sensitivity and specificity \cite{Gulshan2016Development,Gargeya2017Automated}. However, prediction alone does not resolve clinical interpretability. In clinical reasoning, a DR grade is supported by visible lesion types, lesion burden, lesion location, and vascular morphology. Microaneurysms, haemorrhages, exudates, cotton-wool spots and neovascularisation support disease grading, while vessel calibre, density, tortuosity, fractal structure, CRAE, CRVE and AVR describe the retinal vascular bed \cite{Zhang2024RetinalImagingBiomarkers}.

This creates a gap between image-level classifiers and clinically meaningful explanation. Grad-CAM-style methods can highlight class-discriminative image regions \cite{Selvaraju2017GradCAM}, but heatmaps alone do not quantify whether evidence is macular, vessel-adjacent, artery-related, vein-related, or associated with retinal biomarkers. Conversely, tabular oculomics models can expose associations between retinal morphology and disease severity, but they remove the spatial arrangement of lesions and vessels \cite{Chew2025RetinalBiomarkers}. Automated tools such as AutoMorph now make retinal vascular biomarker extraction possible at scale \cite{Zhou2022AutoMorph}, while lesion-aware representation learning provides a way to encode disease-related visual structure more directly \cite{Huang2021LesionBased}. 

We propose a dual-edge spatial--Jacobian image graph for interpretable DR grading. Each fundus image is treated as a graph node. The first edge family, $E^{12}$, captures spatial vessel--lesion similarity after combining $X_1$ vessel maps with $X_2$ lesion evidence. The second edge family, $E^{34}$, captures embedding--biomarker response geometry by mapping $X_3$ lesion-based image embeddings to $X_4$ AutoMorph biomarkers and summarising the local Jacobian. A lightweight two-token attention module fuses both edge systems into a final representation for DR grading and graph-level interpretation. The overall workflow is summarised in Fig.~\ref{fig:workflow}, where the spatial $X_{12}$ and Jacobian $X_{34}$ branches are constructed separately before attention-based fusion. The contribution is therefore not another image classifier alone, but a graph-centred formulation in which the edges themselves have interpretable meaning. 

\section{Materials and Methods}

\subsection{Dataset and evidence streams}
The internal cohort is derived from the APTOS 2019 Blindness Detection dataset \cite{APTOS2019}. We use a carefully curated, non-augmented subset of 2,910 images. Each retained image has four aligned streams: vessel information ($X_1$), lesion evidence maps/statistics ($X_2$), lesion-based contrastive embedding ($X_3$), and AutoMorph morphometric biomarkers ($X_4$). The data are split using a stratified 60/20/20 protocol into 1,746 training, 582 validation, and 582 test images. The stream order is fixed to avoid ambiguity between lesion evidence and image embeddings. $X_1$ captures AutoMorph-derived artery, vein, vessel and macular-region information. $X_2$ contains DR-XAI-style lesion evidence maps derived using Grad-CAM-style localisation \cite{Selvaraju2017GradCAM}. $X_3$ is the 128-dimensional lesion-based contrastive image embedding. A Huang-style \cite{Huang2021LesionBased} lesion-based contrastive ResNet50 checkpoint is used; the released backbone produces 2048-dimensional features, which are projected deterministically into the canonical 128-dimensional schema. $X_4$ contains biomarkers, including density, tortuosity, average width, fractal dimension, CRAE, CRVE, and AVR, which are features derived from images through the AutoMorph package \cite{Zhou2022AutoMorph}. 

\subsection{Spatial vessel--lesion branch}
The spatial branch combines $X_1$ and $X_2$ in image space. Let $A_i$, $R_i$ and $V_i$ denote artery, vein, and vessel maps, and let $L_i$ denote lesion evidence. Instead of flattening these signals independently, spatial interaction channels are constructed:
\begin{equation}
M^{12}_i = \{A_i,R_i,V_i,L_i,L_i\odot A_i,L_i\odot R_i,L_i\odot V_i,Z_i\},
\end{equation}
where $Z_i$ denotes macular-zone information and $\odot$ denotes pixel-wise interaction. These channels encode whether lesion evidence overlaps with arteries, veins, general vessels or macular regions. A compact representation $z^{12}_i$ is extracted, and the spatial edge between images is:
\begin{equation}
E^{12}_{ij}=\cos(z^{12}_i,z^{12}_j).
\end{equation}

\subsection{Jacobian embedding--biomarker branch}
The Jacobian branch combines $X_3$ and $X_4$ in vector space. A differentiable mapper links the lesion-based embedding to a reduced biomarker representation:
\begin{equation}
\hat{x}_{4,i}=f_{\theta}(x_{3,i}).
\end{equation}
For each image, the local Jacobian is:
\begin{equation}
J_i=\frac{\partial f_{\theta}(x_{3,i})}{\partial x_{3,i}}.
\end{equation}
The descriptor $z^{34}_i$ concatenates reduced $X_3$, reduced $X_4$, the Frobenius norm of $J_i$, and input/output sensitivity summaries. The Jacobian edge is:
\begin{equation}
E^{34}_{ij}=\cos(z^{34}_i,z^{34}_j).
\end{equation}
This edge measures similarity in embedding--biomarker response geometry. It is used as an interpretable sensitivity descriptor and is not interpreted as causal evidence.

\subsection{Attention fusion and graph interpretation}
The two branch descriptors are projected into a shared space:
\begin{equation}
t^{12}_i=\phi_{12}(z^{12}_i), \qquad t^{34}_i=\phi_{34}(z^{34}_i).
\end{equation}
A two-token attention module learns branch-specific weights:
\begin{equation}
h_i=\alpha^{12}_i t^{12}_i+\alpha^{34}_i t^{34}_i.
\end{equation}
The fused node representation $h_i$ is used for five-class DR grading and binary referable-DR classification. The final graph stores the attention-weighted combination of $E^{12}$ and $E^{34}$ and supports interpretation through graph topology, branch weights, biomarker sensitivities and lesion--biomarker associations. Spearman trend tests and Benjamini--Hochberg false-discovery-rate correction are used for hypothesis-generating analysis \cite{Benjamini1995FDR}.

\section{Results and Discussion}

\begin{table}[t]
\centering
\caption{APTOS-DR 5-class test performance ($n=582$).}
\vspace{-0.3cm}
\label{tab:results}
\scriptsize
\begin{tabular}{lccccc}
\toprule
Stream / model & Acc. & QWK & Macro-F1 & MAE & Adj. Acc.\\
\midrule
$X_1$ vessel & .6409 & .5038 & .2832 & .6186 & .7887\\
$X_2$ lesion evidence & .7388 & .6720 & .4256 & .4244 & .8729\\
$X_3$ Huang LCL & .8110 & .8265 & .5934 & .2749 & .9278\\
$X_4$ biomarkers & .6770 & .5809 & .3553 & .5498 & .8076\\
$X_{12}$ spatial & .7354 & .6890 & .4258 & .4124 & .8832\\
$X_{34}$ Jacobian & .8007 & .8280 & .5826 & .2835 & .9296\\
Full graph & .8076 & .8312 & .5915 & .2749 & .9330\\
\bottomrule
\end{tabular}
\vspace{-.75cm}
\end{table}

Table~\ref{tab:results} reports the five-class test results. The strongest single stream is the Huang-style lesion-based contrastive embedding $X_3$, which reaches 0.8110 accuracy, 0.8265 QWK and 0.5934 macro-F1. The Jacobian branch $X_{34}$ also performs strongly, reaching 0.8007 accuracy and 0.8280 QWK. The full graph obtains 0.8076 accuracy, 0.8312 QWK, 0.5915 macro-F1 and 0.9330 adjacent-grade accuracy. Although the full graph is close to $X_3$ in raw accuracy, it provides a more interpretable structure by linking prediction to spatial vessel--lesion evidence, embedding--biomarker sensitivity and graph topology. For binary referable DR, the full graph obtains 0.9055 accuracy, 0.9711 AUROC, 0.8964 sensitivity, 0.9111 specificity, 0.8786 F1 and 0.9423 AUPRC. This indicates that the representation is particularly strong when used as a referable-screening layer, where the task is clinically meaningful and less sensitive to uncertainty between adjacent DR grades. The interpretation layer suggests that lesion evidence and vascular morphology are connected in measurable ways within the internal cohort. The strongest corrected associations include conservative neovascularisation and haemorrhage evidence linked with vascular calibre and density measures, especially zone-C biomarkers such as CRVE/CRAE and artery density. 

These findings should be interpreted as hypothesis-generating associations rather than causal claims. These results clarify the role of representation quality and can be a good basis for the next-generation framework of representation, causality, and interpretability in the healthcare domain.  For example, Table~\ref{tab:results} demonstrates that the full graph, after lightweight fusion of spatial--Jacobian features, improves interpretability by showing lesion-burden sensitivity across DR grades using FDR-corrected statistical testing. 
This supports the central argument that graph interpretability depends on meaningful stream construction. The spatial branch asks where lesion-like evidence appears relative to vessels and macular zones, while the Jacobian branch asks which retinal biomarker directions are sensitive to lesion/image embeddings. Collectively, these branches provide a compact, explainable representation for DR grading and oculomics-oriented hypothesis generation.

From an epidemiological and causal-learning perspective, the model does not estimate causal effects, but it creates a structured decomposition of retinal evidence. This decomposition can later be connected to glycaemic exposure, blood pressure, lipid profiles, renal markers or genetic risk to test whether retinal biomarkers act as correlates, mediators or modifiers of disease pathways.

\section*{LLM Use Statement}
AI assistance was used for language editing only; all research design, experiments, results and conclusions are the authors' responsibility.

\bibliographystyle{ACM-Reference-Format}
\bibliography{Reference}


\begin{thebibliography}{9}


\ifx \showCODEN    \undefined \def \showCODEN     #1{\unskip}     \fi
\ifx \showISBNx    \undefined \def \showISBNx     #1{\unskip}     \fi
\ifx \showISBNxiii \undefined \def \showISBNxiii  #1{\unskip}     \fi
\ifx \showISSN     \undefined \def \showISSN      #1{\unskip}     \fi
\ifx \showLCCN     \undefined \def \showLCCN      #1{\unskip}     \fi
\ifx \shownote     \undefined \def \shownote      #1{#1}          \fi
\ifx \showarticletitle \undefined \def \showarticletitle #1{#1}   \fi
\ifx \showURL      \undefined \def \showURL       {\relax}        \fi
\providecommand\bibfield[2]{#2}
\providecommand\bibinfo[2]{#2}
\providecommand\natexlab[1]{#1}
\providecommand\showeprint[2][]{arXiv:#2}

\bibitem[{Asia Pacific Tele-Ophthalmology Society}(2019)]%
        {APTOS2019}
\bibfield{author}{\bibinfo{person}{{Asia Pacific Tele-Ophthalmology Society}}.} \bibinfo{year}{2019}\natexlab{}.
\newblock \bibinfo{title}{APTOS 2019 Blindness Detection}.
\newblock \bibinfo{howpublished}{Kaggle competition dataset}.
\newblock
\urldef\tempurl%
\url{https://www.kaggle.com/c/aptos2019-blindness-detection}
\showURL{%
\tempurl}
\newblock
\shownote{Online dataset}.


\bibitem[Benjamini and Hochberg(1995)]%
        {Benjamini1995FDR}
\bibfield{author}{\bibinfo{person}{Yoav Benjamini} {and} \bibinfo{person}{Yosef Hochberg}.} \bibinfo{year}{1995}\natexlab{}.
\newblock \showarticletitle{Controlling the False Discovery Rate: A Practical and Powerful Approach to Multiple Testing}.
\newblock \bibinfo{journal}{\emph{Journal of the Royal Statistical Society: Series B}} \bibinfo{volume}{57}, \bibinfo{number}{1} (\bibinfo{year}{1995}), \bibinfo{pages}{289--300}.
\newblock
\href{https://doi.org/10.1111/j.2517-6161.1995.tb02031.x}{doi:\nolinkurl{10.1111/j.2517-6161.1995.tb02031.x}}


\bibitem[Chew et~al\mbox{.}(2025)]%
        {Chew2025RetinalBiomarkers}
\bibfield{author}{\bibinfo{person}{Emily~Y. Chew}, \bibinfo{person}{Stephen~A. Burns}, \bibinfo{person}{Alison~G. Abraham}, \bibinfo{person}{Michael~F. Bakhoum}, \bibinfo{person}{Joshua~A. Beckman}, \bibinfo{person}{Toco Y.~P. Chui}, \bibinfo{person}{Robert~P. Finger}, \bibinfo{person}{Alejandro~F. Frangi}, \bibinfo{person}{Rebecca~F. Gottesman}, \bibinfo{person}{Maria~B. Grant}, \bibinfo{person}{Henner Hanssen}, \bibinfo{person}{Cecilia~S. Lee}, \bibinfo{person}{Michelle~L. Meyer}, \bibinfo{person}{Damiano Rizzoni}, \bibinfo{person}{Alicja~R. Rudnicka}, \bibinfo{person}{Joel~S. Schuman}, \bibinfo{person}{Sara~B. Seidelmann}, \bibinfo{person}{W.~H.~Wilson Tang}, \bibinfo{person}{B.~B. Adhikari}, \bibinfo{person}{N. Danthi}, \bibinfo{person}{Y. Hong}, \bibinfo{person}{D. Reid}, \bibinfo{person}{G.~L. Shen}, {and} \bibinfo{person}{Y.~S. Oh}.} \bibinfo{year}{2025}\natexlab{}.
\newblock \showarticletitle{Standardization and Clinical Applications of Retinal Imaging Biomarkers for Cardiovascular Disease: A Roadmap from an NHLBI Workshop}.
\newblock \bibinfo{journal}{\emph{Nature Reviews Cardiology}} \bibinfo{volume}{22}, \bibinfo{number}{1} (\bibinfo{year}{2025}), \bibinfo{pages}{47--63}.
\newblock
\href{https://doi.org/10.1038/s41569-024-01060-8}{doi:\nolinkurl{10.1038/s41569-024-01060-8}}


\bibitem[Gargeya and Leng(2017)]%
        {Gargeya2017Automated}
\bibfield{author}{\bibinfo{person}{Rishab Gargeya} {and} \bibinfo{person}{Theodore Leng}.} \bibinfo{year}{2017}\natexlab{}.
\newblock \showarticletitle{Automated Identification of Diabetic Retinopathy Using Deep Learning}.
\newblock \bibinfo{journal}{\emph{Ophthalmology}} \bibinfo{volume}{124}, \bibinfo{number}{7} (\bibinfo{year}{2017}), \bibinfo{pages}{962--969}.
\newblock
\href{https://doi.org/10.1016/j.ophtha.2017.02.008}{doi:\nolinkurl{10.1016/j.ophtha.2017.02.008}}


\bibitem[Gulshan et~al\mbox{.}(2016)]%
        {Gulshan2016Development}
\bibfield{author}{\bibinfo{person}{Varun Gulshan}, \bibinfo{person}{Lily Peng}, \bibinfo{person}{Marc Coram}, \bibinfo{person}{Martin~C. Stumpe}, \bibinfo{person}{Derek Wu}, \bibinfo{person}{Arunachalam Narayanaswamy}, \bibinfo{person}{Subhashini Venugopalan}, \bibinfo{person}{Kasumi Widner}, \bibinfo{person}{Tom Madams}, \bibinfo{person}{Jorge Cuadros}, \bibinfo{person}{Ramasamy Kim}, \bibinfo{person}{Rajiv Raman}, \bibinfo{person}{Philip~C. Nelson}, \bibinfo{person}{Jessica~L. Mega}, {and} \bibinfo{person}{Dale~R. Webster}.} \bibinfo{year}{2016}\natexlab{}.
\newblock \showarticletitle{Development and Validation of a Deep Learning Algorithm for Detection of Diabetic Retinopathy in Retinal Fundus Photographs}.
\newblock \bibinfo{journal}{\emph{JAMA}} \bibinfo{volume}{316}, \bibinfo{number}{22} (\bibinfo{year}{2016}), \bibinfo{pages}{2402--2410}.
\newblock
\href{https://doi.org/10.1001/jama.2016.17216}{doi:\nolinkurl{10.1001/jama.2016.17216}}


\bibitem[Huang et~al\mbox{.}(2021)]%
        {Huang2021LesionBased}
\bibfield{author}{\bibinfo{person}{Yijin Huang}, \bibinfo{person}{Li Lin}, \bibinfo{person}{Pujin Cheng}, \bibinfo{person}{Junyan Lyu}, {and} \bibinfo{person}{Xiaoying Tang}.} \bibinfo{year}{2021}\natexlab{}.
\newblock \showarticletitle{Lesion-Based Contrastive Learning for Diabetic Retinopathy Grading from Fundus Images}. In \bibinfo{booktitle}{\emph{Medical Image Computing and Computer Assisted Intervention -- MICCAI 2021}} \emph{(\bibinfo{series}{Lecture Notes in Computer Science}, Vol.~\bibinfo{volume}{12902})}. \bibinfo{publisher}{Springer}, \bibinfo{pages}{113--123}.
\newblock
\href{https://doi.org/10.1007/978-3-030-87196-3_11}{doi:\nolinkurl{10.1007/978-3-030-87196-3_11}}


\bibitem[Selvaraju et~al\mbox{.}(2017)]%
        {Selvaraju2017GradCAM}
\bibfield{author}{\bibinfo{person}{Ramprasaath~R. Selvaraju}, \bibinfo{person}{Michael Cogswell}, \bibinfo{person}{Abhishek Das}, \bibinfo{person}{Ramakrishna Vedantam}, \bibinfo{person}{Devi Parikh}, {and} \bibinfo{person}{Dhruv Batra}.} \bibinfo{year}{2017}\natexlab{}.
\newblock \showarticletitle{Grad-CAM: Visual Explanations from Deep Networks via Gradient-Based Localization}. In \bibinfo{booktitle}{\emph{Proceedings of the IEEE International Conference on Computer Vision}}. \bibinfo{pages}{618--626}.
\newblock
\href{https://doi.org/10.1109/ICCV.2017.74}{doi:\nolinkurl{10.1109/ICCV.2017.74}}


\bibitem[Zhang et~al\mbox{.}(2024)]%
        {Zhang2024RetinalImagingBiomarkers}
\bibfield{author}{\bibinfo{person}{Zhengwei Zhang}, \bibinfo{person}{Callie Deng}, {and} \bibinfo{person}{Yannis~M. Paulus}.} \bibinfo{year}{2024}\natexlab{}.
\newblock \showarticletitle{Advances in Structural and Functional Retinal Imaging and Biomarkers for Early Detection of Diabetic Retinopathy}.
\newblock \bibinfo{journal}{\emph{Biomedicines}} \bibinfo{volume}{12}, \bibinfo{number}{7} (\bibinfo{year}{2024}), \bibinfo{pages}{1405}.
\newblock
\href{https://doi.org/10.3390/biomedicines12071405}{doi:\nolinkurl{10.3390/biomedicines12071405}}


\bibitem[Zhou et~al\mbox{.}(2022)]%
        {Zhou2022AutoMorph}
\bibfield{author}{\bibinfo{person}{Yukun Zhou}, \bibinfo{person}{Siegfried~K. Wagner}, \bibinfo{person}{Mark~A. Chia}, \bibinfo{person}{An Zhao}, \bibinfo{person}{Peter Woodward-Court}, \bibinfo{person}{Moucheng Xu}, \bibinfo{person}{Robbert~R. Struyven}, \bibinfo{person}{Daniel~C. Alexander}, {and} \bibinfo{person}{Pearse~A. Keane}.} \bibinfo{year}{2022}\natexlab{}.
\newblock \showarticletitle{AutoMorph: Automated Retinal Vascular Morphology Quantification Via a Deep Learning Pipeline}.
\newblock \bibinfo{journal}{\emph{Translational Vision Science \& Technology}} \bibinfo{volume}{11}, \bibinfo{number}{7} (\bibinfo{year}{2022}), \bibinfo{pages}{12}.
\newblock
\href{https://doi.org/10.1167/tvst.11.7.12}{doi:\nolinkurl{10.1167/tvst.11.7.12}}


\end{thebibliography}

\end{document}